\documentclass[aps,pra,amsmath,amssymb,twocolumn]{revtex4}
\usepackage[english]{babel}
\usepackage{latexsym}
\usepackage{graphics}
\usepackage{subfigure}
\usepackage{epsfig}
\begin{document}
%\wideabs{

\title{Scissors modes of two-component degenerate gases:\\ 
Bose-Bose and Bose-Fermi mixtures}
\author{M. Rodr\'\i guez$^{1}$, P. Pedri$^{2,3}$, 
P. T\"orm\"a$^{4}$ and L. Santos$^2$}
\affiliation{(1) Laboratory of Computational Engineering, P.O. Box 9203, 
FIN-02015 Helsinki University of Technology, Finland}
\affiliation{(2) Institut f\"ur Theoretische Physik, 
Universit\"at Hannover, D-30167 Hannover,Germany}
\affiliation{(3) Dipartimento di Fisica, Universit\`a di Trento 
and BEC-INFM, I-38050 Povo, Italy}
\affiliation{(4) Department of Physics, University of Jyv\"askyl\"a, P.O.Box 
35, FIN-40014 Jyv\"askyl\"a, Finland}

\begin{abstract}
We investigate the scissors modes in binary mixtures of degenerate 
dilute quantum gases, for both Bose-Bose and Bose-Fermi mixtures. 
For the latter we consider both the 
superfluid and normal hydrodynamic and collisionless regimes. 
We analyze the dependence of 
the frequencies of the scissors modes and their character 
as a function of the 
Bose-Fermi coupling and the trap geometry.  We show that the scissors mode 
can reveal a clear trace of the hydrodynamic behavior of the Fermi gas.
\end{abstract}
\pacs{03.75.Mn,03.75.Ss}
\maketitle

\section{Introduction}

% Bose-bose mixtures.

In recent years, the improvement of the trapping techniques 
has allowed for the creation of multi-component Bose-Einstein condensates (BECs), formed 
by atoms in different internal (electronic) states \cite{JILA2,MIT}. 
The multicomponent BEC,   
far from being a trivial extension of the single-component one, presents novel and  
fundamentally different scenarios for its ground-state  
wavefunction \cite{Esry1,Lo,Patrik1} and excitations \cite{Thomas2,Esry2,Patrik2}. 
In particular, it has been experimentally observed that the BEC can reach   
an equilibrium state characterized by the phase separation of the species in 
different domains \cite{MIT}. 

% Fermi-Bose mixtures.

During the last few years, the mixtures of fermions and bosons, from now on 
called Fermi-Bose (FB) mixtures, have also attracted a growing attention. 
Although the initial interest has been mostly motivated by the 
possibility to achieve sympathetic cooling towards the 
Bardeen-Cooper-Schriffer (BCS) transition \cite{revBCS,bfcooling}, 
recently, the rich physics of FB mixtures 
has become itself 
one of the central topics of the physics of ultracold gases. 
Various properties of the FB mixtures have been  
analyzed, including the phase separation between bosons 
and fermions \cite{molmer,stoof,pu}, the existence of novel types of 
collective modes \cite{stoof,pu,capuzzi,liu}, the appearance of effective 
Fermi-Fermi interactions mediated by the bosons \cite{stoof,pu,albus,viverit}, the  
collapse of the Fermi cloud in the presence of attractive interactions 
between bosons and fermions \cite{collapse}, 
the physics of 1D FB mixtures \cite{1DBF} or FB mixtures 
in optical lattices \cite{BFLatt}.

% Collective excitations in general 

The analysis of collective excitations is an excellent tool to analyze 
the effects of interactions in ultracold dilute gases. 
Zero-temperature excitation frequencies have been extensively studied in
the case of Bose-Einstein condensates (BECs) 
in dilute alkali gases up to a high degree of 
experimental accuracy \cite{JILAexc,MITexc}. These experimental results 
are in excellent agreement with the theoretical predictions obtained 
from first principles \cite{exc_th,Perez,tom}.

% Scissors
Among the different collective modes the so-called scissors mode has attracted 
considerable attention. This mode is well known in nuclear physics \cite{iudice}, where 
it corresponds to the out-of-phase rotation of the neutron and proton clouds. 
In the context of ultracold atomic gases, the scissors mode 
is achieved after a sudden tilt of the trap, 
and it consists in the harmonic motion of the atomic cloud 
around the new trap axis. 
This particular collective excitation,  
predicted in Ref.~\cite{stringari1}, has been successfully 
observed experimentally \cite{foot,Modugno}.

% What are we going to do and motivation.

The present paper is devoted to the analysis of the scissors modes in 
mixtures of degenerate atomic gases, both for the case of a multicomponent 
BEC, and for a FB mixture. In the latter case, we analyze both when the 
Fermi gas is in hydrodynamic and in collisionless regime. 
We discuss in particular the variation of the frequencies 
of the scissors modes (as well as their associated geometry)
as a function of the Bose-Bose (or Bose-Fermi) 
coupling constant.

% Scheme of the paper

The paper is organized as follows. In Sec.~\ref{sec:BB} we analyze the case 
of a Bose-Bose mixture. Sec.~\ref{sec:BF} is devoted to the analysis of the 
scissors modes in FB mixtures, first for the case of a hydrodynamic  
Fermi gas, and later when the Fermi gas is in the normal phase. 
We finalize in Sec.~\ref{sec:conclu} with our conclusions, and the discussion of 
the results. In App.~\ref{app:A} we have included some analytical 
results, which have been employed in different calculations throughout the paper.

\section{Binary Bose-Einstein Condensate}
\label{sec:BB}

In this section we consider the case of a 
BEC formed by two different components $\{ 1,2 \}$, 
confined in a harmonic external 
potential
\begin{equation}
V_i({\bf r})=\frac{m_i}{2}\left [  \omega_{x;i}^2 x^2+
\omega_{y;i}^2 y^2+\omega_{z;i}^2 z^2
\right ],
\end{equation}
where $i=1,2$.
We are interested in the scissors modes of this particular system after a 
sudden tilt of the trap axis. 
We assume that the chemical potential is larger than the energy separation between the 
trap levels (Thomas-Fermi regime). 
Therefore the effects of the quantum pressure are negligible, and the system 
can be analyzed using the hydrodynamic equations for superfluids  
\begin{eqnarray}
&&\frac{\partial n_1}{\partial t}+\nabla\cdot(n_1{\bf v}_1)=0\\
&&\frac{\partial n_2}{\partial t}+\nabla\cdot(n_2{\bf v}_2)=0 \\
&&m_1\frac{\partial {\bf v}_1}{\partial t}+\nabla
\left(
\frac{m_1}{2}v_1^2+V_1+g_{11}n_1+g_{12}n_2
\right)=0\\
&&m_2\frac{\partial {\bf v}_2}{\partial t}+\nabla
\left(
\frac{m_2}{2}v_2^2+V_2+g_{12}n_1+g_{22}n_2
\right)=0
\end{eqnarray}
where $n_i$ is the density, ${\bf v}_i$ the velocity field, $m_i$ the mass, 
and $V_i$ the external potential, associated with the $i$-th component. The coefficient 
$g_{ij}=2\pi\hbar^2a_{ij}/{\tilde m}_{ij}$ is the coupling constant associated with the 
interatomic interactions between the components $i$ and $j$, where ${\tilde m}_{ij}$ 
and $a_{ij}$, are the corresponding reduced mass and scattering length, respectively.

%%%%%%%%%%%%%%%%%%%%%%
% FIGURE 1
\begin{figure}[ht!]
\begin{center}\
\psfig{file=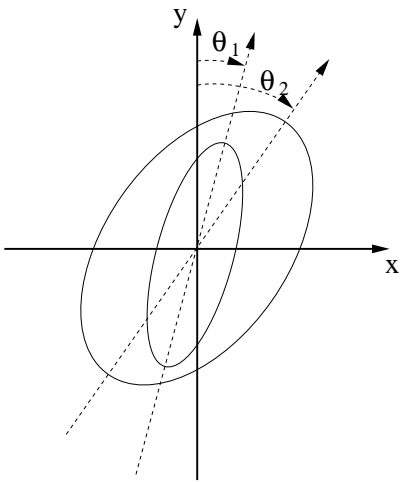,width=5.3cm}\\[0.1cm]
\end{center} 
\caption{Schematic description of the two angles of rotation involved in 
the scissors modes.}
\label{density}  
\end{figure}
%%%%%%%%%%%%%%%%%

Neglecting the quantum pressure term in a single component BEC demands 
that $Na/l_{ho}\gg1$, 
where $N$ is the total number of particles 
and $l_{ho}=\sqrt{\hbar/m\bar\omega}$, where 
$\bar\omega^3=\omega_x\omega_y\omega_z$. This 
approximation breaks down at the border of the Thomas-Fermi density profile, 
where the density is small. In the case of a 
two-component BEC, one must additionally require 
that the two species are not strongly phase separated. 
Otherwise, at the domain wall, 
the kinetic energy dominates the mean-field potential, and therefore the quantum pressure 
is certainly not negligible.

Imposing a stationary solution without flow (${\bf v=0}$), we obtain
\begin{eqnarray}
&&V_1({\bf r})+g_{11}n_{1 g}({\bf r})+g_{12}n_{2 g}({\bf r})
=\mu_1
\label{TF1}
\\
&&V_2({\bf r})+g_{12}n_{1 g}({\bf r})+g_{22}n_{2 g}({\bf r})=\mu_2, 
\label{TF2}
\end{eqnarray}
from where we can easily calculate the ground-state density profiles 
$n_{i g}({\bf r})$.

Since we are interested in the oscillations of the cloud in the $x$-$y$ plane after a 
tilting from the equilibrium solution, it is convenient to adopt the coordinates 
$x'_i=x\cos(\theta_i(t))-y\sin(\theta_i(t)))$ and 
$y'_i=x\sin(\theta_i(t))-y\cos(\theta_i(t)))$. 
We consider small angles $\theta_i$, and consequently we can perform the Ansatz
$n_i(t,x,y,z)=n_{i g}(x'_i,y'_i,z)\simeq n_{g}(x,y,z)+
\theta_i(y\partial_x-x\partial_y)n_{gx}(x,y,z)$. Note that the time dependence 
is entirely contained in the parameters $\theta_1$ and $\theta_2$ (see Fig. 1).

After multiplying the continuity equations by $xy$ and integrating, we obtain 
the equation for the asymmetry of the cloud of the $i$-th component
\begin{equation}
\frac{d}{dt}\langle xy\rangle_i=\langle xv_y+yv_x\rangle_i. \label{conti}
\end{equation}
Note that the asymmetry parameter is related to the angular displacement $\theta_i$ in the 
form $\langle xy \rangle_i=\langle y^2-x^2 \rangle_{g} \theta_i$, where 
$\langle\dots\rangle_{g}$ denotes the expected value in the equilibrium state.
Neglecting terms of second order in the angular displacements, we obtain
\begin{equation}
\frac{d}{dt}\langle xv_y+yv_x\rangle
=\frac{1}{N}\int (y,x,0)\cdot
\left[\frac{\partial{\bf v}}{\partial t}\right]nd^3r.
\label{kk}
\end{equation}
Substituting into Eq.~(\ref{kk}) the corresponding Euler equation for the 
$i$-th components, and combining with Eqs. (\ref{conti}), we obtain, 
in the linear regime  
\begin{eqnarray}
&&B_{11}\ddot{\theta}_1=A_{11}\theta_1+A_{12}\theta_2\label{B1}\\
&&B_{22}\ddot{\theta}_2=A_{21}\theta_1+A_{22}\theta_2\label{B2},
\end{eqnarray} 
where the coefficients $A_{ij}$ and $B_{i}$ can be found in App.~\ref{app:A}. 
By imposing a solution of the form $\exp\{-i\omega t\}$, we find the 
frequencies of the different scissors modes
\begin{equation}
\omega_{\pm}=\frac{A_{11}B_2+A_{22}B_1\pm\sqrt{(A_{11}B_2+A_{22}B_1)^2
-{\rm det [{\bf A}]}}}{B_1B_2}, \label{wpm}
\end{equation}
where we have introduced the matrix ${\bf A}=\{A_{ij}\}$. 
The corresponding eigenvectors $\{\theta_1^{\pm},\theta_2^{\pm}\}$ provide information 
about the nature of the modes. The two components oscillate in-phase when 
$\theta_1\theta_2>0$, and in counter-phase, 
when $\theta_1\theta_2<0$. 

In Fig. 2 we depict our results for a particular set of parameters (see figure caption). 
Both in-phase and out-of-phase modes depend 
on the particular value of  $g_{12}$. Notice, however, 
that the in-phase mode is less affected by the change of $g_{12}$, since the 
interaction energy between both gases is just slightly modified during the motion.
In particular, if the trapping frequencies for both components are the same, 
the in-phase mode remains independent of the 
interaction coupling constant $g_{12}$.   
The counter-phase mode is in any case strongly affected. 

\begin{figure}[ht!]
\begin{center}\
\psfig{file=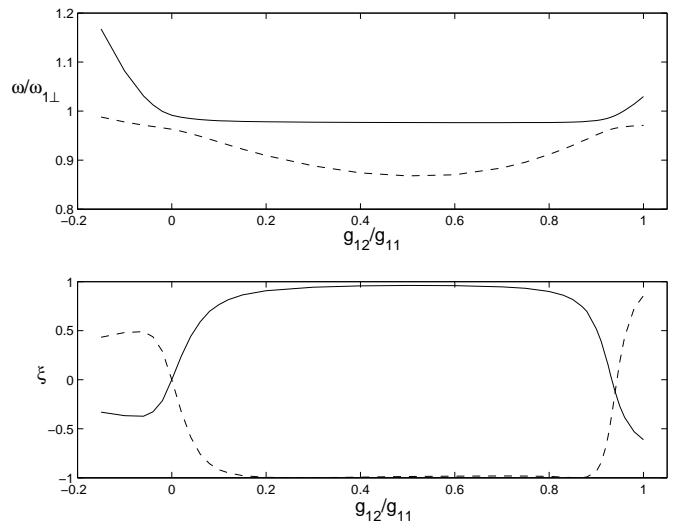,width=8.8cm}\\[0.1cm]
\end{center} 
\caption{(a) Frequencies of the scissors modes for a binary BEC as a function of the 
intercomponent coupling constant $g_{12}$ for $N_1=N_2=10^6$, 
$\omega_{x,y,z;2}=0.97\omega_{x,y,z;1}$, equal masses $m$, 
$\omega_{y;1}=1.5\omega_{x;1}$, and $\omega_{z;1}=2\omega_{x;1}$. 
The dimensionless 
interaction energies are $g_{11}/\hbar\omega_{x;1}l_{x;1}^3=1$ and $g_{22}=2g_{11}$,
where $l_{x;1}^2=\hbar/m\omega_{x;1}$.
The frequencies are indicated in units of $\omega_{1\perp}=(\omega_{x;1}^2+\omega_{y;1}^2)^{1/2}$.
(b) Character of the modes, 
provided by $\xi=2\theta_1\theta_2/(\theta_1^2+\theta_2^2)$, as a function of 
$g_{12}$. The in-phase (out-of-phase) mode is characterized by $\xi>0$ ($\xi<0$).
}
\label{bosebose}  
\end{figure}

The in-phase mode is always the lowest energy state for $g_{12}<0$, since 
it maximizes the overlapping. For $g_{12}>0$, the out-of-phase mode 
becomes the state with lowest energy, as long as the two components 
present a maximum in the trap center, since in that case the out-of-phase mode 
minimizes the overlapping. For a given value $g_{12}=g_{12}^{cr}$, 
the density profile of the component 
with lowest central density (say the component $2$) 
acquires zero curvature at the trap center, therefore, 
for $g_{12}>g_{12}^{cr}$, one of the components presents a minimum at the trap center, 
and  the in-phase mode becomes once more the lowest energy state. 
For $\omega_{x,y,z;1}=\omega_{x,y,z;2}$, $g_{12}^{cr}=g_{11}$, whereas from the 
Thomas-Fermi solutions (\ref{TF1}) and 
(\ref{TF2}), one can easily 
obtain that in general, if $\beta=\omega_{x,y,z;1}/\omega_{x,y,z;2}$, 
$g_{12}^{cr}/g_{11}=\beta^2 (m_1/m_2)$. 
This dependence of the modes has been confirmed by our numerical simulations 
(see Fig. \ref{bosebose}). 

\section{Bose-Fermi mixture}
\label{sec:BF}

In this section we analyze the scissors modes for FB mixtures. 
Two different regimes are considered:
(i) The case in which the fermions are in the hydrodynamic regime, and 
(ii) the case in which the fermions are in the normal phase in the collisionless regime. 
Concerning the case (i), we would like to note that the scissors modes 
for a FB mixture in the normal phase in the hydrodynamic regime, 
coincide with those modes of the superfluid hydrodynamics. 
In Sec. IV we comment on this issue in more detail.

\subsection{Hydrodynamic regime}

In a two-component Fermi gas, if the typical collisional time becomes much larger than 
the trapping period, the Fermi gas is said to enter into the hydrodynamic regime.  
As for the Bose-Bose mixture, the FB 
mixture in the hydrodynamic regime can be well described by means of the 
hydrodynamic equations
\begin{eqnarray}
&&\frac{\partial n_b}{\partial t}+\nabla\cdot(n_b{\bf v}_b)=0 \label{contb}\\
&&\frac{\partial n_f}{\partial t}+\nabla\cdot(n_f{\bf v}_f)=0 \label{contf} \\
\frac{\partial {\bf v}_b}{\partial t}&+&\nabla
\left(
\frac{v_b^2}{2}+\frac{V_b+g_bn_b+g_{bf}n_f}{m_b}
\right)=0 \label{eulerb} \\
\frac{\partial {\bf v}_f}{\partial t} &+& \nabla
\left(
\frac{v_f^2}{2}+\frac{V_f+g_{bf}n_b}{m_f}+\frac{\hbar^2}{2m_f^2}(6\pi^2n_f)^{2/3}
\right) = \nonumber \\
&=& {\bf v}_f\wedge ({\bf\nabla}\wedge{\bf v}_f). \label{eulerf}
\end{eqnarray}  
Note that in our calculations only one set of equations is considered for the 
Fermi gas. This is justified if the difference between 
the trapping frequencies and the
concentrations of both Fermi components is negligible \cite{Vichi}.
Note also that we have employed the chemical potential for a free Fermi gas, 
$\sim n_f^{2/3}$, since in the quantum degeneracy regime due to Pauli-blocking 
(away from Feshbach resonances) the corrections 
due to Fermi-Fermi interactions are small.

%%%%%%%%%%%%%%%%%%%%%%
% FIGURE 2
%%%%%%%%%%%%%%%%%

%%%%%%%%%%%%%%%%%%%%%%
% FIGURE 3
\begin{figure}[ht!]
\begin{center}\
\psfig{file=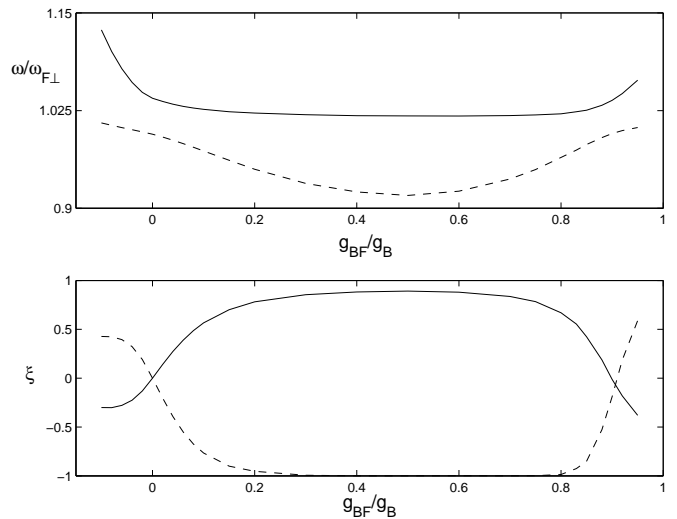,width=8.8cm}\\[0.1cm]
\end{center} 
\caption{(a) Frequencies of the scissors modes for a FB mixture, where the Fermi gas 
is in the hydrodynamic regime, 
in units of $\omega_{F\perp}=(\omega_{x;f}^2+\omega_{y;f}^2)^{1/2}$, 
as a function of the 
intercomponent coupling constant $g_{bf}$; (b) Character of the modes, 
provided by $\xi=2\theta_b\theta_f/(\theta_b^2+\theta_f^2)$, as a function of 
$g_{bf}$. The in-phase (out-of-phase) mode is characterized by $\xi>0$ ($\xi<0$). 
In the figures we consider the case $N_b=N_f=10^6$, equal masses $m$, 
$\omega_{x,y,z;b}=1.05\omega_{x,y,z;f}$, 
$\omega_{x;f}=0.1\omega_{z;f}$, $\omega_{y;f}=0.995\omega_{z;f}$, 
and $g_{b}/\hbar\omega_{z;f}l_{z;f}^3=1$.
}
\label{fermiboseh}  
\end{figure}
%%%%%%%%%%%%%%%%%

As for the Bose-Bose mixture, we obtain the ground state density profile 
in the Thomas-Fermi 
approximation, and introduce two time dependent parameters $\theta_b$ and $\theta_f$, 
which describe the small angular displacements 
of the bosonic and fermionic cloud, respectively. 
From the hydrodynamic equations we obtain the equations of motion for the angles
\begin{eqnarray}
&&C_1\ddot{\theta}_b=D_{11}\theta_b+D_{12}\theta_f \label{C1} \\
&&C_2\ddot{\theta}_f=D_{21}\theta_b+D_{22}\theta_f \label{C2}. 
\end{eqnarray}   
Although the Eqs.(\ref{C1}) and (\ref{C2}) are formally the 
same as Eqs. (\ref{B1}) and (\ref{B2}), the coefficients are obviously
different. The form of these coefficients is detailed in App.{\ref{app:A}. 
The frequencies of the scissors modes are obtained in a similar way 
as in Eqs. (\ref{wpm}). A typical result is depicted in Fig.~\ref{fermiboseh}. 
Note that similarly to the case of Bose-Bose mixtures, and due to similar reasons, 
the out-of-phase mode becomes the less energetic state for $0<g_{bf}<g_{bf}^c$.

Note that, contrary to the Bose-Bose case, Eq.~(\ref{eulerf}) allows for the 
possibility of rotational flow, which could be present if the Fermi gas 
is in the hydrodynamic regime, but still in the normal phase. However, 
since we are only interested in small deviations with respect to equilibrium, 
the rotational term, as well as the kinetic energy term, are neglected, since 
they provide corrections of second order in the angular displacements.

%%%%%%%%%%%%%%%%%%%%%%
% FIGURE 4
\begin{figure}[ht!]
\begin{center}\
\psfig{file=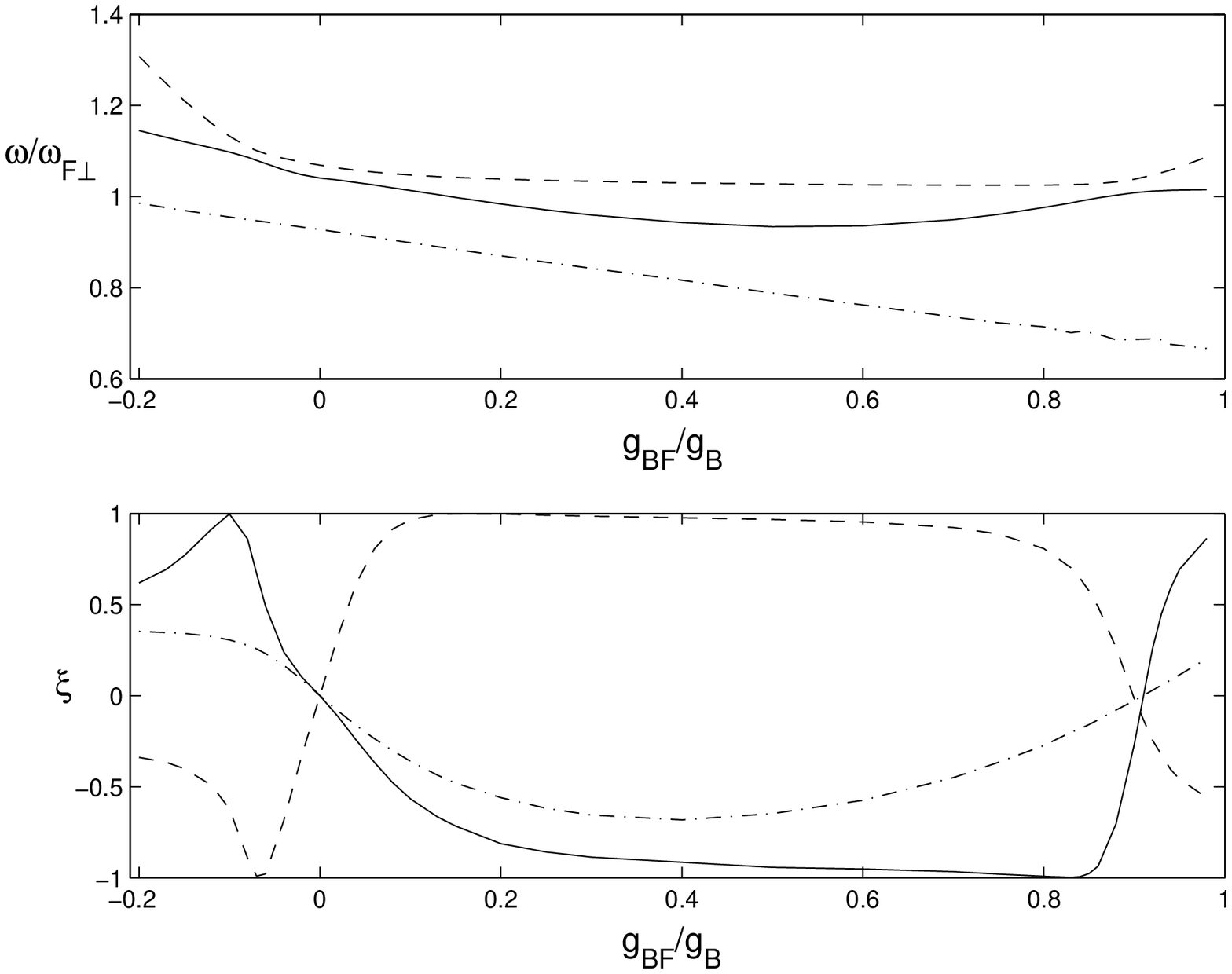,width=8.8cm}\\[0.1cm]
\end{center} 
\caption{(a) Frequencies of the scissors modes for a FB mixture, where the Fermi gas 
is in the collisionless regime, , 
in units of $\omega_{F\perp}=(\omega_{x;f}^2+\omega_{y;f}^2)^{1/2}$, as a function of the 
intercomponent coupling constant $g_{bf}$; (b) Character of the modes, 
provided by $\xi=2\theta_b\theta_f/(\theta_b^2+\theta_f^2)$, as 
a function of $g_{bf}$. The in-phase (out-of-phase) mode is characterized 
by $\xi>0$ ($\xi<0$). Same parameters as in Figure \ref{fermiboseh} are considered.
}
\label{fermibosec}  
\end{figure}
%%%%%%%%%%%%%%%%%

\subsection{Collisionless regime}

If the typical time for Fermi-Fermi collisions becomes much smaller 
that the trapping period, 
the system enters in the collisionless regime. In this case 
the bosons still follow the superfluid hydrodynamic equations (\ref{contb}) 
and (\ref{eulerb}), but the fermionic gas 
follows the so-called Boltzmann-Vlasov equation \cite{Huang}
\begin{equation}
\frac{\partial f}{\partial t}+{\bf v}\cdot\nabla_{\bf r}f+
\frac{1}{m_f}\nabla_{\bf r}U_f\cdot\nabla_{\bf v}f=0,
\end{equation}
where $f$ is the phase-space distribution function for the fermions. 
The external potential, $U_f({\bf r})=V_f({\bf r})+g_{bf}n_b({\bf r})$, contains both 
the harmonic confinement for the fermions ($V_f$) 
and the mean field term due to the 
interaction with the bosons. The fermionic density is defined as 
$n_f(t,{\bf r})=\int d^3vf(t,{\bf r},{\bf v})$.

Combining the hydrodynamic equations for the bosons and the Boltzmann-Vlasov 
equation for the fermions, we obtain the following set of equations
\begin{eqnarray}
&&\hspace*{-0.9cm}\frac{d}{dt}\langle xy \rangle_b = \langle xv_y+yv_x \rangle_b 
\label{kk1}\\
&&\hspace*{-0.9cm}\frac{d}{dt}\langle xv_y+yv_x \rangle_b = \nonumber \\
&&\hspace*{-0.9cm}= \left\langle \left[ \left (x\partial_y+y\partial_x \right )
\left(\frac{m_b}{2}v^2_b+V_b+g_bn_b+g_{bf}n_f \right ) \right ] \right \rangle_b\\
&&\hspace*{-0.9cm}\frac{d}{dt}\langle xy \rangle_f=\langle xv_y \rangle_f+\langle yv_x \rangle_f\\
&&\hspace*{-0.9cm}\frac{d}{dt}\langle xv_y \rangle_f=\langle v_xv_y \rangle_f+
\frac{1}{m_f}\langle x[\partial_y(V_f+g_{bf}n_b)]\rangle_f\\
&&\hspace*{-0.9cm}\frac{d}{dt}\langle yv_x \rangle_f=\langle v_xv_y \rangle_f+
\frac{1}{m_f}\langle y[\partial_x(V_f+g_{bf}n_b)]\rangle_f\\
&&\hspace*{-0.9cm}\frac{d}{dt}\langle v_xv_y \rangle_f=\langle
[(v_x\partial_y+v_y\partial_x)(V_f+g_{bf}n_b)]\rangle_f,
\label{kk2}
\end{eqnarray}
where $V_b$ is the harmonic confinement for the bosons, 
$\langle \chi\rangle_b = \int \chi({\bf r}) n_b(t,{\bf r}) d^3r/N_b $, and 
$\langle \chi \rangle_f = \int \chi({\bf r},{\bf v}) f(t,{\bf r},{\bf v}) d^3rd^3v/N_f$.

After a rather tedious algebraic manipulation we arrive to 
the following equations of motion for the 
displacement angles (see App. {\ref{app:A}):
\begin{eqnarray}
L_1\ddot{\theta}_b&=&M_{11}\theta_b+M_{12}\theta_f \label{thetab}\\
L_2\ddot{\theta}_f&=&M_{21}\theta_b+M_{22}\theta_f+M_{23}\eta \label{thetaf}\\
\ddot{\eta}&=&M_{31}\theta_b+M_{32}\theta_f+M_{33}\eta \label{eta}.
\end{eqnarray}
Note that in the collisionless regime a third variable couples with the 
displacement angles, namely 
$\eta=\int v_xv_y f d^3rd^3v/N_f$, 
which is the average value of the deformation in momentum
space \cite{stringari1}. Note that in the hydrodynamic regime $\eta=0$, since in the 
hydrodynamic case the momentum distribution remains always spherically symmetric.
A typical case is depicted in Fig.~\ref{fermibosec}. Note the 
appearance of three scissors modes, 
and the mixing of the bosonic and the collisional fermionic modes.

\section{Conclusions}
\label{sec:conclu}

% Conclusions

In this paper we have analyzed the scissors modes obtained after a sudden 
tilting of the trap axis for the case of binary mixtures of quantum gases.
We have studied the scissors modes for the case of 
a BEC with two components, and especially the dependence of 
the modes with the intercomponent coupling constant. Whereas for 
attractive intercomponent interactions the in-phase mode always presents the lowest 
energy, we have shown that for the case of repulsive interactions 
the character of the modes depends on the density 
profile of the ground state, and in particular on the overlapping of both components. 
If both components present the maximal density at the trap center, the 
lowest energy corresponds to the out-of-phase mode. At a given critical 
coupling constant one of the components develops a minima at the center, and the 
in-phase mode becomes the less energetic. A similar reason explains 
the behavior of other collective excitations in binary mixtures \cite{liu}. 

In the second part of the paper we have focused our attention on the case 
of a mixture of degenerated bosonic and fermionic gases. We have 
first analyzed the case in which the fermionic gas is in the hydrodynamic regime, 
in which the scissors modes present a similar behavior as for the case of 
binary condensates. We have completed our analysis with the detailed study of 
the case in which the Fermi gas is in the normal collisionless phase. 
In that case, the system presents three different scissors modes, due to the 
asymmetry of the velocity distribution of the fermionic component.
 
Finally, we would like to stress that
 the difference between the superfluid hydrodynamic regime and 
the collisional hydrodynamic regime 
is provided, from the macroscopic point of view,  
by the irrotational constraint for the velocity field in the superfluid 
case ($\nabla\wedge{\bf v}=0$), 
related to the quantization of the flux and the appearance of quantized vortices. 
On the contrary, this condition is not necessarily 
fulfilled in the collisional hydrodynamics. As discussed above,  
the scissors mode is excited after a sudden change of the trap axis. 
In the absence  of friction such a sudden change cannot induce a 
rotational flow. Consequently, the scissors 
modes in collisional hydrodynamics do not differ from those expected in the 
superfluid case. The differences between collisional 
and superfluid hydrodynamics in FB mixtures 
can be revealed extending to the FB mixtures similar ideas as those discussed 
for Fermi gases in Ref.~\cite{cozzini}. 
This problem will be the subject of further studies.

{\em Note added} After the completion of this work, we have learned of a related 
work performed by Kasamatsu {\it et al.} \cite{Kasamatsu} 
on the problem of scissors modes in binary 
condensates, in which the authors employ different techniques as those discussed in this paper. 
In Ref.~\cite{Kasamatsu} special emphasis is paid to the nonlinear coupling between 
quadrupole and scissors modes.

\begin{acknowledgments}
\end{acknowledgments}
We acknowledge  support from Deutsche Forschungsgemeinschaft (SFB 407),  
the RTN Cold Quantum gases, ESF PESC BEC2000+, and the Ministero dell'Istruzione, 
dell'Universit\`a e della Ricerca (MIUR). M. R. acknowledges the 
hospitality of the Hannover group.
L. S. and P. P. wish to thank the Alexander von Humboldt Foundation, 
the Federal Ministry of Education and 
Research and the ZIP Programme of the German Government. 
M. R. and P. T. thank the Academy of Finland (Project Nos.\ 53903, 48445) for support.\\

\begin{appendix}
\section{}
\label{app:A}

The coefficients of Eqs.~(\ref{B1}) and (\ref{B2}) can be obtained after a somewhat 
tedious calculation following the method described in the text, and acquire the form:
\begin{eqnarray}
B_1&=&\frac{1}{N_1}\int xyn'_{1g}d^3r\\
B_2&=&\frac{1}{N_2}\int xyn'_{2g}d^3r\\
A_{11}&=&\frac{g_{11}}{N_1}\int n_{1g}(y\partial_x+x\partial_y)n'_{1g}d^3r\\
A_{12}&=&\frac{g_{12}}{N_1}\int n_{1g}(y\partial_x+x\partial_y)n'_{2g}d^3r\\
A_{21}&=&\frac{g_{12}}{N_2}\int n_{2g}(y\partial_x+x\partial_y)n'_{1g}d^3r\\
A_{22}&=&\frac{g_{22}}{N_2}\int n_{2g}(y\partial_x+x\partial_y)n'_{2g}d^3r,
\end{eqnarray}
where $n_{ig}$ denotes the equilibrium density, and 
$n'_{ig}=(y\partial_x-x\partial_y)n_{ig}$.

Similarly, we can obtain the coefficients of Eqs.~(\ref{C1}) and (\ref{C2}):
\begin{eqnarray}
C_1&=&\frac{1}{N_b}\int xyn'_{1g}d^3r\\
C_2&=&\frac{1}{N_f}\int xyn'_{2g}d^3r\\
D_{11}&=&\frac{g_{b}}{N_b}\int n_{bg}(y\partial_x+x\partial_y)n'_{bg}d^3r\\
D_{12}&=&\frac{g_{bf}}{N_b}\int n_{bg}(y\partial_x+x\partial_y)n'_{fg}d^3r\\
D_{21}&=&\frac{g_{bf}}{N_f}\int n_{fg}(y\partial_x+x\partial_y)n'_{bg}d^3r\\
D_{22}&=&\frac{g_{ff}}{N_f}\int n_{fg}(y\partial_x+x\partial_y)
\frac{2 n'_{fg}}{3n_{fg}^{1/3}} d^3r,
\end{eqnarray}
where $g_{ff}=\hbar^2(6\pi)^{2/3}/2m_f$. 

Finally, let us briefly discuss the derivation of Eqs.~(\ref{thetab}), (\ref{thetaf}) and 
(\ref{eta}). We assume a velocity field for the fermions of the form
\begin{eqnarray}
v_x(t,{\bf r})=\alpha_1y\\
v_y(t,{\bf r})=\alpha_2x
\end{eqnarray}
where the time dependence is contained in the parameters $\alpha_1$ and $\alpha_2$. We 
additionally define
\begin{eqnarray}
\langle v_x v_y \rangle&=&\eta\\
\langle xv_y+yv_x \rangle &=& \beta. 
\end{eqnarray}
It is possible to rewrite Eqs.~(\ref{kk1})-(\ref{kk2}) in the form
\begin{eqnarray}
E_1\dot{\theta}_b&=&\beta\\
\dot{\beta}&=&F_1\theta_b+F_2\theta_f\\
(G_2-G_1)\dot{\theta}_f&=&G_1\alpha_1+G_2\alpha_2\\
G_1\dot{\alpha}_1&=&\eta+Q_{11}\theta_b+Q_{12}\theta_f\\
G_2\dot{\alpha}_2&=&\eta+Q_{21}\theta_b+Q_{22}\theta_f\\
\dot{\eta}&=&R_1\alpha_1+R_2\alpha_2
\end{eqnarray}
where
\begin{eqnarray}
E_1&=&\frac{1}{N_b}\int xyn'_{bg}d^3r\\
F_1&=&\frac{g_b}{N_b}\int n_{bg}(y\partial_x+x\partial_y)n'_{bg}d^3r\\
F_2&=&\frac{g_{bf}}{N_b}\int n_{bg}(y\partial_x+x\partial_y)n'_{fg}d^3r\\
G_1&=&\frac{1}{N_f}\int y^2n_{fg}d^3r \\
G_2&=&\frac{1}{N_f}\int x^2n_{fg}d^3r \\
Q_{12}&=& \int \left [ \frac{\omega_x^2}{N_f} xyn'_{fg} +
\frac{g_{bf}}{N_f}n'_{fg}y\partial_xn_{bg} \right ]  d^3r 
\end{eqnarray}
\begin{eqnarray}
Q_{11}&=&\frac{g_{bf}}{N_f}\int n_{fg}y\partial_xn'_{bg}d^3r \\
Q_{22}&=&\int \left [
\frac{\omega_y^2}{N_f}xyn'_{fg} +
\frac{g_{bf}}{N_f} n'_{fg}x\partial_yn_{bg} \right ] d^3r\\
Q_{21}&=&\frac{g_{bf}}{N_f}\int n_{fg}x\partial_yn'_{bg}d^3r\\
R_1&=&\omega_y^2G_1+\frac{g_{bf}}{N_f}\int n_{bg}y\partial_yn_{bg}d^3r\\
R_2&=&\omega_x^2G_2+\frac{g_{bf}}{N_f}\int n_{bg}x\partial_xn_{bg}d^3r
\end{eqnarray}

The coefficients of Eqs.~(\ref{thetab}-\ref{eta}) are of the form
\begin{eqnarray}
L_1&=&E_1 \\
L_2&=&G_2-G_1 \\
M_{11}&=&F_1\\
M_{22}&=&F_2\\
M_{21}&=&Q_{11}+Q_{21}\\
M_{22}&=&Q_{12}+Q_{22}\\
M_{23}&=&2\\
M_{31}&=&\frac{R_1Q_{11}}{G_1}+\frac{R_2Q_{21}}{G_2}\\
M_{32}&=&\frac{R_1Q_{12}}{G_1}+\frac{R_2Q_{22}}{G_2}\\
M_{33}&=&\frac{R_1}{G_1}+\frac{R_2}{G_2}
\end{eqnarray}

\end{appendix}

\end{document}